# An improved point-to-surface contact algorithm with penalty method for Peridynamics

Haoran Zhang[a,b], Jun Li[a,b], Hai Mei[a,b], Xin Lai[a,b,*], Lisheng Liu[a,b,**]

[a] Hubei Longzhong Laboratory, Wuhan University of Technology Xiangyang Demonstration Zone, Xiangyang 441000, China

[b] Hubei Key Laboratory of Theory and Application of Advanced Materials Mechanics, Wuhan University of Technology, Wuhan 430070, China

[*]Corresponding Author: Xin Lai. Email: laixin@whut.edu.cn

[**]Corresponding Author: Lisheng Liu. Email: liulish@whut.edu.cn

## Abstract

In Peridynamics (PD), it is challenging to identify surface particles and determine contact forces accurately for complex contact and collision problems. In the present study, we propose an improved point-to-surface contact algorithm with penalty function method for PD simulations. This contact algorithm enables the automatic identification of surface particles, which thus provides a straightforward and accurate estimation of contact forces. The algorithm first employs the eigenvalue method to detect external surface particles, followed by a global search using Verlet lists to identify all potential contact particle pairs. Subsequently, a local point-to-surface contact search is employed to precisely determine contact locations. Based on the identified surface particles and their exact contact positions, the contact forces can be calculated via the penalty function method. Finally, to demonstrate the accuracy of the proposed contact algorithm, several typical contact problems are performed and analyzed, including contact interactions between an elastic body and a rigid plane as well as two elastic bodies. The calculated contact forces exhibit better agreement with those derived from Hertz contact theory than those from other PD contact models. This study introduces a robust and precise contact algorithm for PD simulations, offering valuable insights for investigating complex multi-body interactions and contact phenomena in practical applications.

## **1.** Introduction

Peridynamics (PD), a non-local theory of solid mechanics, was first proposed by Silling in 2000 [1–3]. It uses integral equations to formulate mechanical problems[4], in stark contrast to partial differential equations in classical continuum mechanics theory. And the PD also contains a new definition of damage and failure [5], which thus exhibits spontaneous capability to depict the initiation and propagation of cracks. Therefore, the PD method is adept at examining the mechanical behaviors of structures with cracks or other discontinuities [6].

Generally, impact, extrusion, friction, and closure crack problems often involve complex contact issues [7] with triple nonlinearities in geometry, material behavior, and boundary conditions. Since it is challenging to capture internal deformations and failure behaviors within contacting structures experimentally, numerical simulation has become a critical way to address contact problems. To date, based on different discrete methods, contact algorithms can be classified into three categories: node-to-node/particle-to-particle (NTN/PTP) [8,9], node-to-segment/particle-to-surface (NTS/PTS) [10], and segment-to-segment/surface-to-surface (STS/STS) [11]. For example, in the finite element method (FEM) [12–15], Xing et al.[16] proposed a NTN contact scheme within the Scaled Boundary Finite Element Method framework for three-dimensional frictional contact problems, which formulates three-dimensional (3D) Coulomb friction via MCP in complementarity form, thereby obtaining accurate solutions for small-deformation elastostatic contact. Sun et al. [17] proposed an NTS contact strategy combining the node-based smooth finite element method and the edge-based smooth finite element method for the two-dimensional (2D) frictionless contact problem. Mayer et al.[18] proposed a STS contact element approach for joint interfaces in FEM, employing zero-thickness/thin-layer elements originally from geomechanics. The method eliminates global contact searches for fixed joints with small relative motions, significantly improving computational efficiency. While for meshless methods, Libersky [19] first applied the smoothed particle hydrodynamics (SPH) method to hypervelocity impact simulations, which defined boundaries as surfaces offset by half the smoothing length from the boundary SPH particles. Some contact algorithms were then proposed, such as penalty function methods for PTP interactions[20], hybrid PTP/PTS algorithms[21], and three-dimensional STS contact algorithms combining surface reconstruction with detection criteria[22].

In addition, some contact algorithms were also proposed for contact problems in PD modeling. For example, Madenci introduced a contact-collision algorithm to describe the contact behaviors between rigid impactor and deformable target [23] , in which the material points inside the impactor are relocated to their new positions outside the impactor after impact. This algorithm has been widely used to describe contact collision problems in PD simulations [24–26]. But it is only applicable to the contact problem when the impactor is a rigid body and the target is a deformable body. While for deformable impactors (e.g., metal), Macek and Silling[27] proposed a novel contact algorithm, where a short-range force was defined to prevent material particles from penetrating each other [28–31]. However, this algorithm can only identify the interactions of material particles rather than the contact surfaces, so it is difficult to determine the clear contact boundary. Moreover, this method does not consider the important normal force and friction on the contact surface. Kamensky et al[32]. considered the effect of friction by applying a pair of frictional forces perpendicular to the bond direction between the particles of the contacting material. Although this method takes into account the friction effect, the direction of the friction force is difficult to be controlled. Thus, the model has not been addressed the contact force ambiguity issues caused by the lack of determined contact surfaces. Silling[33] proposed a bond-based friction-adhesion model, which applies additional force pairs parallel to the bond of the predetermined relative slip surfaces to represent the friction effect. But this model requires to determine the contact surfaces first, which complicates the prediction and identification of the contact surfaces. Overall, the current contact models without considering the contact surface cannot accurately analyze the normal forces, while the models considering the contact surface exhibit the difficulty of identifying the contact surface[32–36]. Therefore, it is necessary to develop an improved contact algorithm for PD simulations, which can automatically identify the outer surface of the model and then determine the contact force accurately.

In the present study, inspired by the determination of free-surface particles in SPH and the NTS contact algorithms in FEM, we propose an improved point-to-surface contact algorithm with penalty method for PD modeling. First, the surface particles are identified using the eigenvalue method [37] based on the assumption that contact only occurs on the model's outer surface. A Verlet list is then constructed for searching potential contact pairs among outer surface particles. Next, we propose a point-to-surface contact search algorithm to determine the exact positions of contact pairs, and then calculate the contact forces using the penalty method. Finally, the accuracy of the proposed contact algorithm is demonstrated by comparing contact forces obtained from this contact mode, classical Hertz contact theory, and other PD contact methods.

## 2. Methodology

In this section, we will introduce the proposed improved point-to-surface contact algorithm with the penalty method.

### *2.1. Basic Equations of PD*

The PD theory is a nonlocal theory proposed by Silling [1–3], which utilizes spatial integral equations to describe the deformations, especially for discontinuous bodies. It can be primarily divided into two categories: bond-based PD (BB-PD) [1,5] and state-based PD (SB-PD) [2,3]. The BB-PD model[1,5] assumes that any two points within the horizon are connected with a bond through spring-like interaction. While for the state-based peridynamics (SB-PD) [2,3] the bond force density of the points depends on its whole family deformation.

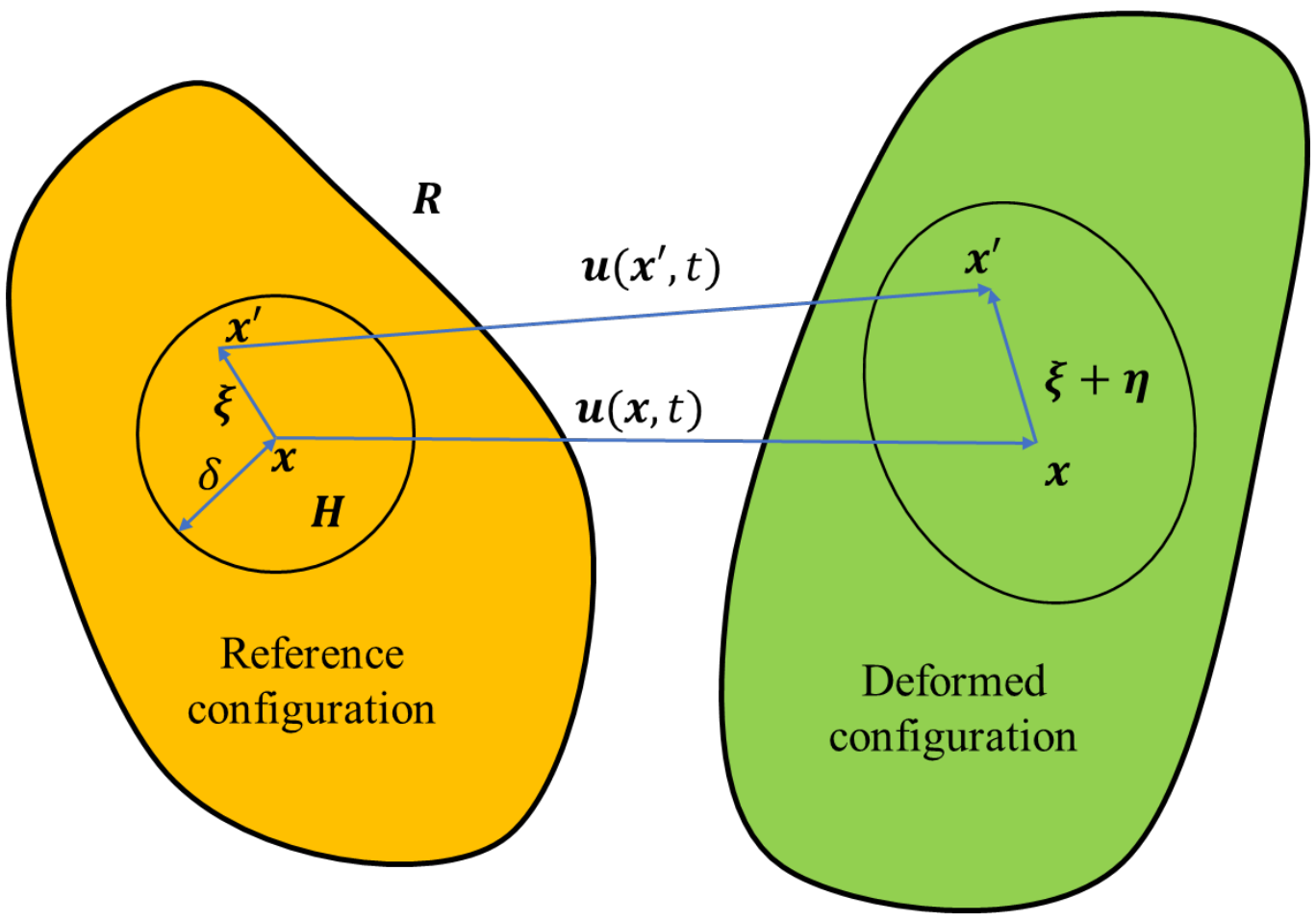

**Fig.1.** Illustration of the bond-based peridynamic model.

Noteworthy, the proposed contact algorithm can be applicable to both BB-PD and SB-PD simulations, however, we mainly employed the BB-PD to illustrate the accuracy of the proposed contact method in the present study due to computational efficiency. Thus, we only briefly introduce the BB-PD theory here. In BB-PD theory, a material body occupies a spatial domain which is composed of a series of discrete particles and their interacting bonds. As shown in Fig.1. At any time t, the motion equation of particle $\boldsymbol{x}$ in the BB-PD can be written as:

$$\rho\ddot{\boldsymbol{u}}(\boldsymbol{x},\mathrm{t}) = \int_{\boldsymbol{H}_x} \boldsymbol{f}(\boldsymbol{\xi},\boldsymbol{\eta})\,\mathrm{d}V_{x'} + \boldsymbol{b}(\boldsymbol{x},\mathrm{t}) \tag{1}$$

where $\rho$ denotes the mass density, $\ddot{\boldsymbol{u}}$ is the acceleration vector field of the particle $\boldsymbol{x}$, $\boldsymbol{b}$ represents the body force density of the particle $\boldsymbol{x}$, $\boldsymbol{H}_x$ denotes the horizon of the particle $\boldsymbol{x}$ in the spatial domain. The pairwise bond force vector is a function of the initial relative position vector $\boldsymbol{\xi}$ in the reference configuration and the relative displacement vector $\boldsymbol{\eta}$ in the current configuration, as shown in Eq. (2):

$$\boldsymbol{\xi} = \boldsymbol{x}' - \boldsymbol{x}, \boldsymbol{\eta} = \boldsymbol{u}(\boldsymbol{x}',\mathrm{t}) - \boldsymbol{u}(\boldsymbol{x},\mathrm{t}) \tag{2}$$

The pairwise bond force vector $\boldsymbol{f}$ for a homogeneous objective microelastic material can be written as[5]

$$\boldsymbol{f}(\boldsymbol{\xi},\boldsymbol{\eta}) = cs\boldsymbol{n} \tag{3}$$

where $\boldsymbol{f}(\boldsymbol{\xi},\boldsymbol{\eta})$ is the bond force vector function, $\boldsymbol{n} = (\boldsymbol{\xi} + \boldsymbol{\eta})/|\boldsymbol{\xi} + \boldsymbol{\eta}|$ is the normal unit vector of the bond pair after deformation, $s$ is the relative stretch of the bond, and $c$ is the microelastic modulus of the object [5]. The expressions of $s$ and $c$ are shown in Eq. (4)-(5), respectively:

$$s = \frac{|\eta+\xi|-|\xi|}{|\eta+\xi|} \tag{4}$$

$$c = \frac{6E}{\pi\delta^4(1\text{-}2\nu)}, \nu=\frac{1}{4} \tag{5}$$

where $E$ represents the elastic modulus of the material, $\delta$ represents the size of the neighborhood horizon, and $\nu$ represents the Poisson's ratio of the material.

The motion equation of bond-based PD under spatial dispersion is shown as follows:

$$\rho_i \ddot{\boldsymbol{u}}_i^n = \sum_j \boldsymbol{f}\left(\boldsymbol{u}_j^n - \boldsymbol{u}_i^n, \boldsymbol{x}_j - \boldsymbol{x}_i\right) V_j + \boldsymbol{b}_i^n \tag{6}$$

where $n$ is the iteration time step, subscript $i$, $j$ is the number of material point, $V_j$ denotes the volume occupied by material point $j$, $\boldsymbol{f}$ is the bond pair force, which can be expressed as:

$$\boldsymbol{f}\left(\boldsymbol{u}_j^n - \boldsymbol{u}_i^n, \boldsymbol{x}_j - \boldsymbol{x}_i\right) = c s_{ij}^n \boldsymbol{n}_{ij}^{\boldsymbol{n}} \tag{7}$$

where $s_{ij}^n$ represents the relative elongation between material point $i$ and $j$ bond pairs at the $\mathrm{n}th$ time step, and $\boldsymbol{n}_{ij}^n$ is the unit vector between material point $i$ and $j$ bond pairs at the $\mathrm{n}th$ time step.

### *2.2. The point-to-surface contact algorithm for PD*

Generally, the calculation of contact force for contact-collision problems mainly involves two aspects: contact search and contact force estimation. Here, contact search is divided into two steps: global search and local search, where the former is to find potential contact pairs effectively and the latter is to further accurately identify the real contact state of the contact pairs based on the global search. The contact pairs identified during local search were employed to calculate the contact force.

#### 2.2.1. *Searching for Surface Particles*

Since the surface cannot be determined directly in PD theory, in the present study, we first employed the eigenvalue method [37] to identify particles on the surface and calculate the normal vector of particles.

First, we defined a renormalization matrix $\boldsymbol{B}$ [19] , as shown in Eq. (8):

$$\boldsymbol{B}^{-1} = \sum_{\boldsymbol{H}_x} \nabla \boldsymbol{W} \otimes \boldsymbol{r} V_{\boldsymbol{x}'} \tag{8}$$

where $\boldsymbol{r}$ denotes the relative position vector in the current configuration, and $\boldsymbol{r} = \boldsymbol{\xi} + \boldsymbol{\eta}$, $V_{\boldsymbol{x}'}$ is the volume of the particle $\boldsymbol{x}'$, which is the neighbor of the particle $\boldsymbol{x}$, $\otimes$ indicates the dyadic product, $\nabla\boldsymbol{W}$ denotes the gradient of $\boldsymbol{W}$, and $\boldsymbol{W}$ is the improved Gaussian kernel function [38], as shown in Eq. (9)-(10),

$$\boldsymbol{W} = W(\boldsymbol{r}, \Delta) = \begin{cases} \alpha_d \left[\frac{e^{-(r/\Delta)^2} - e^{-9}}{1-10e^{-9}}\right] & if\ r \leq 3\Delta \\ 0 \quad otherwise \end{cases} \tag{9}$$

$$\alpha_d = \begin{cases} \frac{1}{(\pi^{1/2}\Delta)} & 1\text{-}D \\ \frac{1}{(\pi\Delta^2)} & 2\text{-}D \\ \frac{1}{(\pi^{3/2}\Delta^3)} & 3\text{-}D \end{cases} \tag{10}$$

The value of the minimum eigenvalue $\lambda$ of the matrix $\boldsymbol{B}^{-1}$ depends on the distribution of the particle $\boldsymbol{x}'$ near the considered particle $\boldsymbol{x}$. The value $\lambda$ theoretically tends to 0 when the particle is far from the surface, while $\lambda$ theoretically tends to 1 when inside the body. Therefore, we can identify whether the particle is located at the surface of the model according to the value of the minimum eigenvalue $\lambda$ of the matrix $\boldsymbol{B}^{-1}$. Then the normal vector of the particles on the surface can be calculated by Eq. (11).

$$\boldsymbol{v} = -\boldsymbol{B}\sum_{H_x} \nabla\boldsymbol{W} V_{x'}\,;\ \boldsymbol{n} = \frac{\boldsymbol{v}}{|\boldsymbol{v}|} \tag{11}$$

### 2.2.2. *Contact Neighbor List for Particles*

Based on the identified surface particles, the Verlet list [39,40] is then used to determine the possible locations for all surface particles, as illustrated in Fig.2. With particle $\boldsymbol{x}$ as the center, the potential cutoff sphere with its radius $\delta_{cut} = \Delta$, is formed. A skin is added to the outside of the cutoff sphere with a radius of $\delta_m = 1.3\Delta$. The thickness of the skin is $\delta_l = \delta_m - \delta_{cut} = 0.3\Delta$.

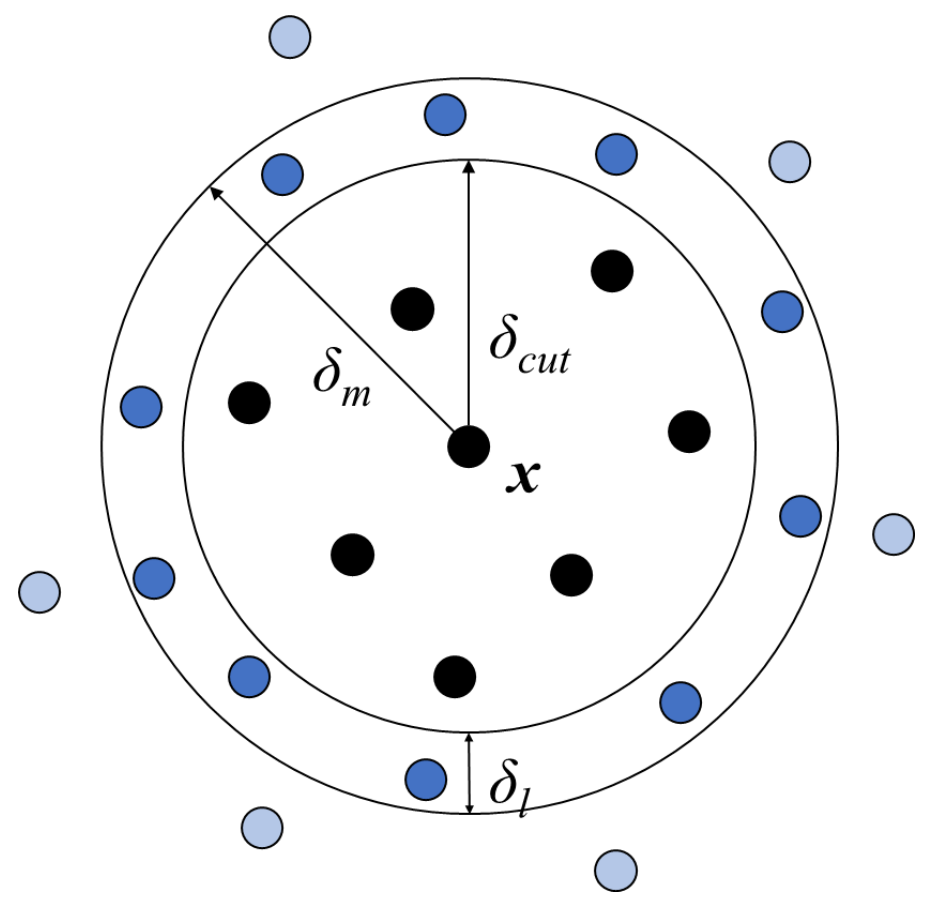

**Fig.2.** The schematic diagram of the cutoff sphere.

The initial Verlet list is constructed by nested loops over all particles in the system, which is completed every $N_m$ time step. The value of $N_m$ satisfies $\delta_m - \delta_{cut} > N_m \bar{v} \Delta t$, where $\bar{v}$ is the particle velocity and $\Delta t$ is the time step. The update interval of the original Verlet list is fixed, regardless of whether the particles move fast enough. This generally wastes much computational time for larger-scale PD calculations. Therefore, it is necessary to refine the Verlet list to allow updates as needed. Here, it is achieved by maintaining a list of the maximum displacement among all particles after each update of the Verlet list [41].The following steps are taken to implement the algorithm:

1. A list $\Delta s_x$ is constructed to store the accumulated displacement vector of particle $\boldsymbol{x}$ after the last Verlet list update;

2. In each time loop, $\Delta s_x$ is updated as $\Delta s_x = \Delta s_x + \Delta u_x$, where $\Delta u_x$ is the displacement of particle $\boldsymbol{x}$ within each time step loop;

3. The two largest values of particle displacement are calculated and denoted as $\Delta s_{max}$ and $\Delta s_{max2}$;

4. Before the end of each time step, the values of $\Delta s_{max}$ and $\Delta s_{max2}$ are compared with the value of $\delta_l$;

5. If $\Delta s_{max} + \Delta s_{max2} > \delta_l$, the Verlet list is updated and the accumulated particle displacement $\Delta s_x$ is reset to zero for the next Verlet neighbor list update. Otherwise, the Verlet list is not updated.

2.2.3. *Point-to-Surface Contact Algorithm*

After constructing the Verlet list, we used the point-to-surface contact algorithm [10,42] to determine the position of each potential contact pair, and thus the precise contact pairs are identified. To improve the computational efficiency, we establish a contact neighbor list for the surface based on the surface particles, because contact can only occur on the outer surface. The list of contact neighbors on the surface can be expressed as:

$$CN_f(A) = CN_n(i) \cup CN_n(j) \cup CN_n(k) \cup CN_n(l) \tag{12}$$

where $CN_f(A)$ is the contact neighbor list of surface $A$; $CN_n(a)$, $(a = i, j, k, l)$ is the contact neighbor list of points that make up the surface $A$.

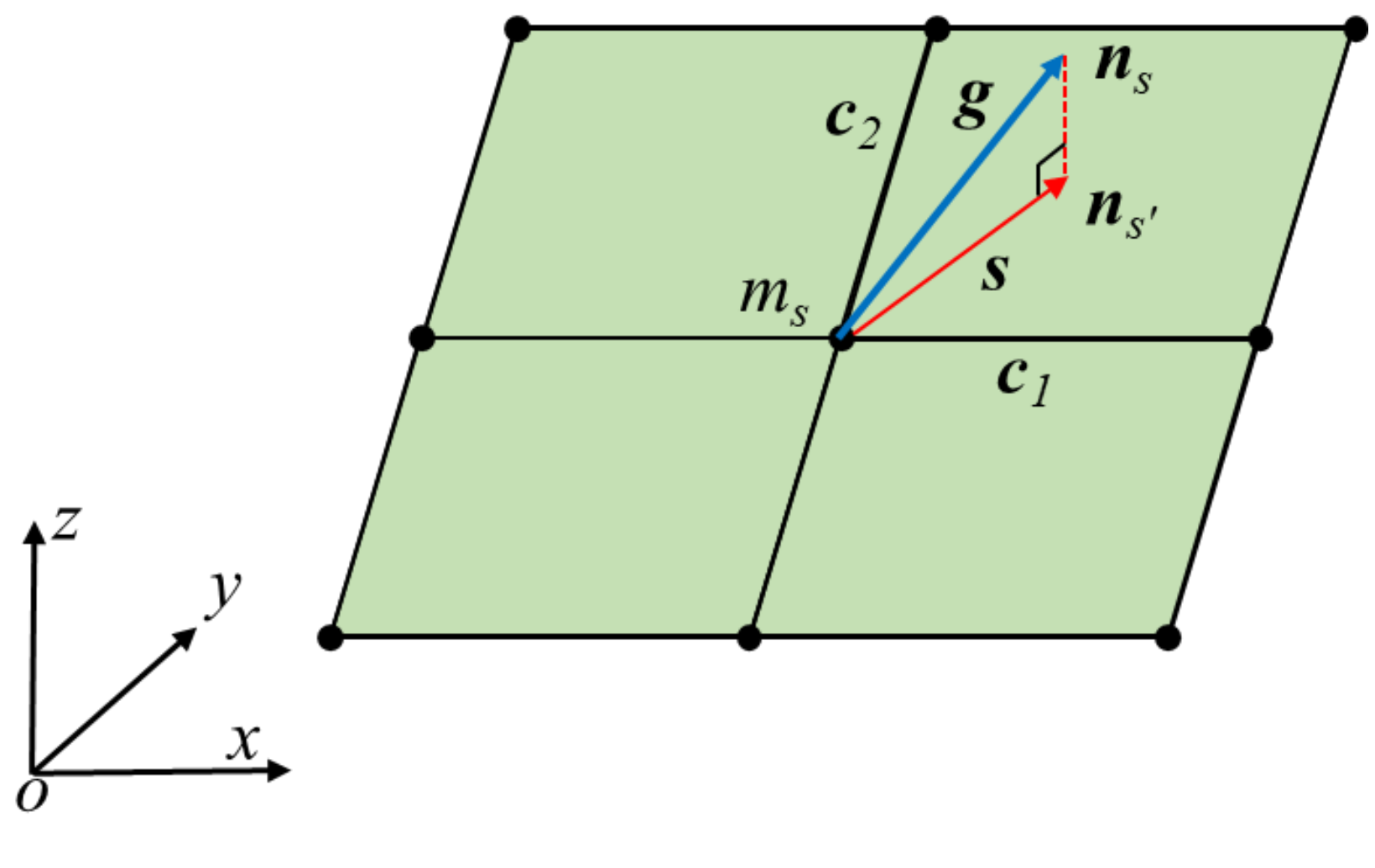

**Fig.3.** The projection of the vector **g** onto the principal surface.

After constructing a list of contact neighbors on the surface, the point-to-surface contact search algorithm is employed to determine the contact state between material points and contact surfaces. As illustrated in Fig.3, the contact condition between a slave point $n_s$ and the main surface $m_s$ is determined as follows: when the projection of the slave point lies within the main surface and the gap distance to the main surface is less than $0.5\Delta$, the slave point is considered as a real contact pair with the main surface in the current time step, and the contact force generated by the slave point is calculated using the penalty function method. Otherwise, there is no contact between the slave point and the main surface in the current time step.

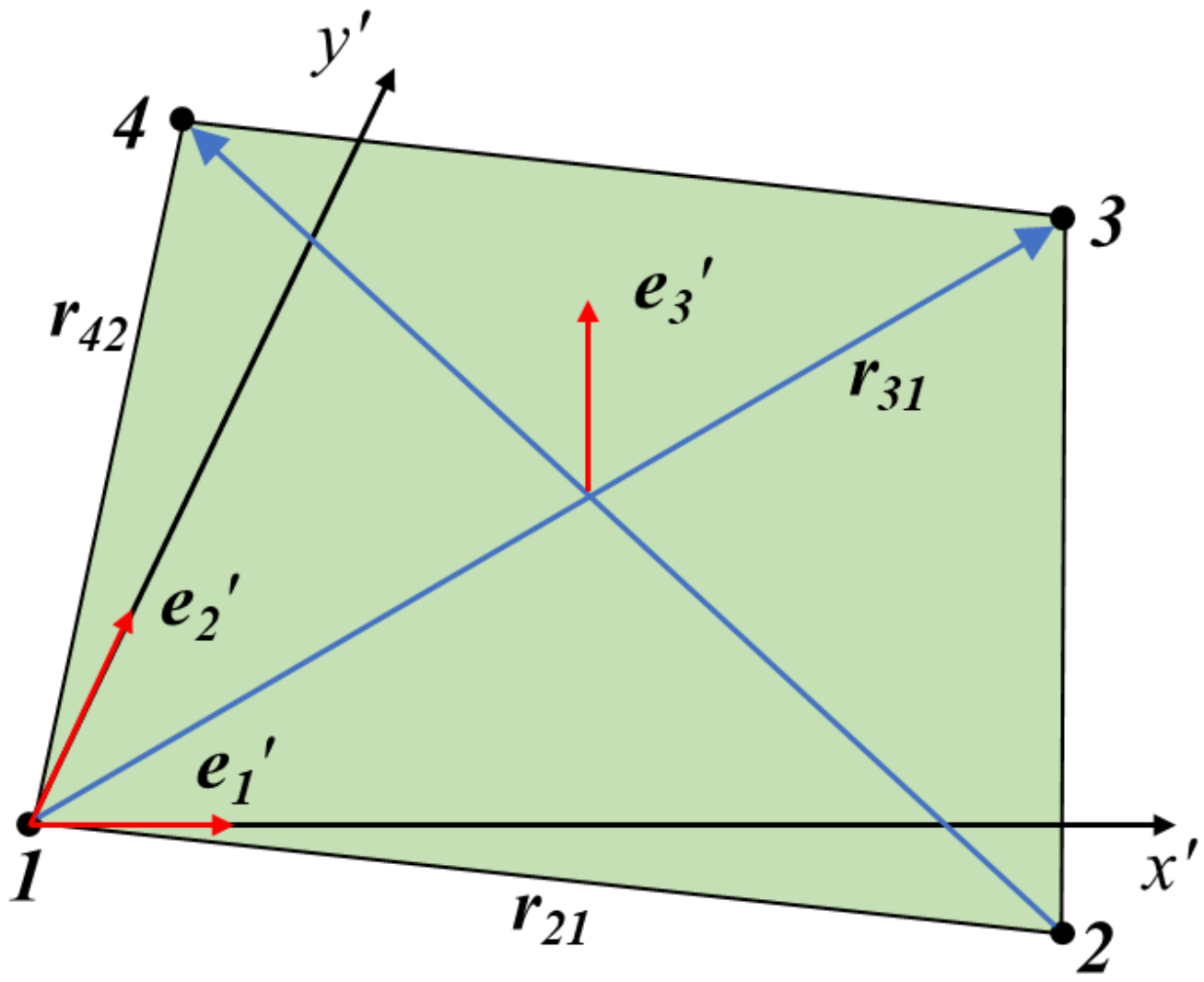

**Fig.4.** The projection plane constructed.

In the calculation of projection points from the slave point to the main surface [43], both the point and the surface are projected onto a plane, as displayed in Fig. 4. Thus, the three-dimensional space problem can be simplified into a two-dimensional plane problem. The base vector of the projection plane can be expressed as:

$$\boldsymbol{e}_3' = \frac{\boldsymbol{r}_{31} \times \boldsymbol{r}_{42}}{\|\boldsymbol{r}_{31} \times \boldsymbol{r}_{42}\|}$$

$$\boldsymbol{e}_1' = \frac{\boldsymbol{r}_{21} - (\boldsymbol{r}_{21} \cdot \boldsymbol{e}_3')\boldsymbol{e}_3'}{\|\boldsymbol{r}_{21} - (\boldsymbol{r}_{21} \cdot \boldsymbol{e}_3')\boldsymbol{e}_3'\|} \tag{13}$$

$$\boldsymbol{e}_2' = \boldsymbol{e}_3' \times \boldsymbol{e}_1'$$

Here, the coordinates of the slave point on the projector plane are $(x', y')$, and the coordinates of the point on the main surface of the projector plane are $(x_i', y_i')$, $(i = 1 \sim 4)$. A bilinear interpolation function is used to describe the main surface, and the local coordinates $(r, t)$ are constructed. If the slave point is in the main surface, the coordinates of the slave point can be expressed as,

$$\begin{cases} x' = \sum_{i=1}^{4} N_i(r, t) x_i' \\ y' = \sum_{i=1}^{4} N_i(r, t) y_i' \end{cases} \tag{14}$$

$$N_i(r, t) = \frac{1}{4}(1 + r_i r)(1 + t_i t) \tag{15}$$

where $(r, t)$ is the local coordinate of the slave point, and $(r_i, t_i)$ is the local coordinate of the point in the main surface. If the calculated local coordinates *r* and *t* of the slave point are in the interval [-1, 1], the projection of the slave point to the main surface is inside the main surface. The distance from the point to the main surface can be calculated by projecting the local coordinates $(r, t)$ from the point onto the main surface:

$$\boldsymbol{\delta_p} = [\boldsymbol{x}^s - \boldsymbol{x}_c(r, t)] \cdot \boldsymbol{n}^m \tag{16}$$

where: $\boldsymbol{x}^s$ is the coordinate of the slave point; $\boldsymbol{x}_c$ is the coordinates of the contact point (projection point); $\boldsymbol{n}^m$ is the normal vector for the primary surface.

The following steps are taken to implement the algorithm:

1. A contact neighbors list is constructed for the surface by using the Verlet list of particles;
2. A projection surface is constructed by using Eq. (13) to transform the three-dimensional space problem into a two-dimensional plane problem;

3. Bilinear interpolation functions are used to describe points on the surface;

4. The particles from the surface contact neighbor list are projected to the projection surface, assuming that the projection point is located inside the main surface, which represents the position of the projection point through an interpolation function, denoted as $(r, t)$;

5. If the calculated local coordinates $r$ and $t$ of the slave point are in the interval [-1, 1], the particle's projection is inside the main surface, and the distance between the point and the main surface is calculated using Eq. (16); Otherwise, the projection of the particle is not within the main surface, and the surface continues to loop to the next particle in the surface contact neighbor list.

2.2.4. *Contact Force*

The penalty function method [44] is used to calculate the contact force between contact pairs. The classical Hertz law relates the contact force to a nonlinear power function of penetration depth, which can be expressed as:

$$F_N = K\boldsymbol{\delta}_p^n \tag{17}$$

where $\boldsymbol{\delta_p}$ represents the relative penetration depth between the contacts, $n$ is the nonlinear power exponents and $K$ is the contact stiffness parameter determined by the material and geometric properties of the local region of the contact body. Here, the contact is considered to be a point-to-surface contact problem between a point and a plane of a substance. The expression for the contact stiffness [45] is:

$$K = \frac{4}{3(m_i + m_j)}\sqrt{R_i} \tag{18}$$

where material parameters $m_i$ and $m_j$ are given by the following formula

$$m_l = \frac{1-v_l^2}{E_l}, (l = i, j) \tag{19}$$

where, $\boldsymbol{v}_l$ and $E_l$ are the Poisson's ratio and elastic modulus of the two contact objects respectively.

## 3. Numerical Implementation

In this section, we will explain the program implementation of the contact algorithm and the typical numerical simulation.

### *3.1. Contact algorithm*

In the present study, we construct the PD model on the basis of the ANSYS finite element model, since it is straightforward to identify and label outer element faces in FEM model. First, both geometric and material information of elements are extracted from ANSYS to the PD model. And the respective faces of these imported elements are determined using eigenvalue method, especially for those located at the boundary. Notably, these element faces do not used in calculating bond forces in PD programs. The Verlet list is then utilized to build a list of positions and perform a global search for potential contact points among these element faces. Next, an eigenvalue method is employed to distinguish different types of element faces based on the minimum eigenvalue. It ensures that only valid external element faces are selected for further analysis. The process is illustrated through a series of flowcharts (Fig.5 and Fig.6), which provide a detailed overview of the identification and classification procedures.

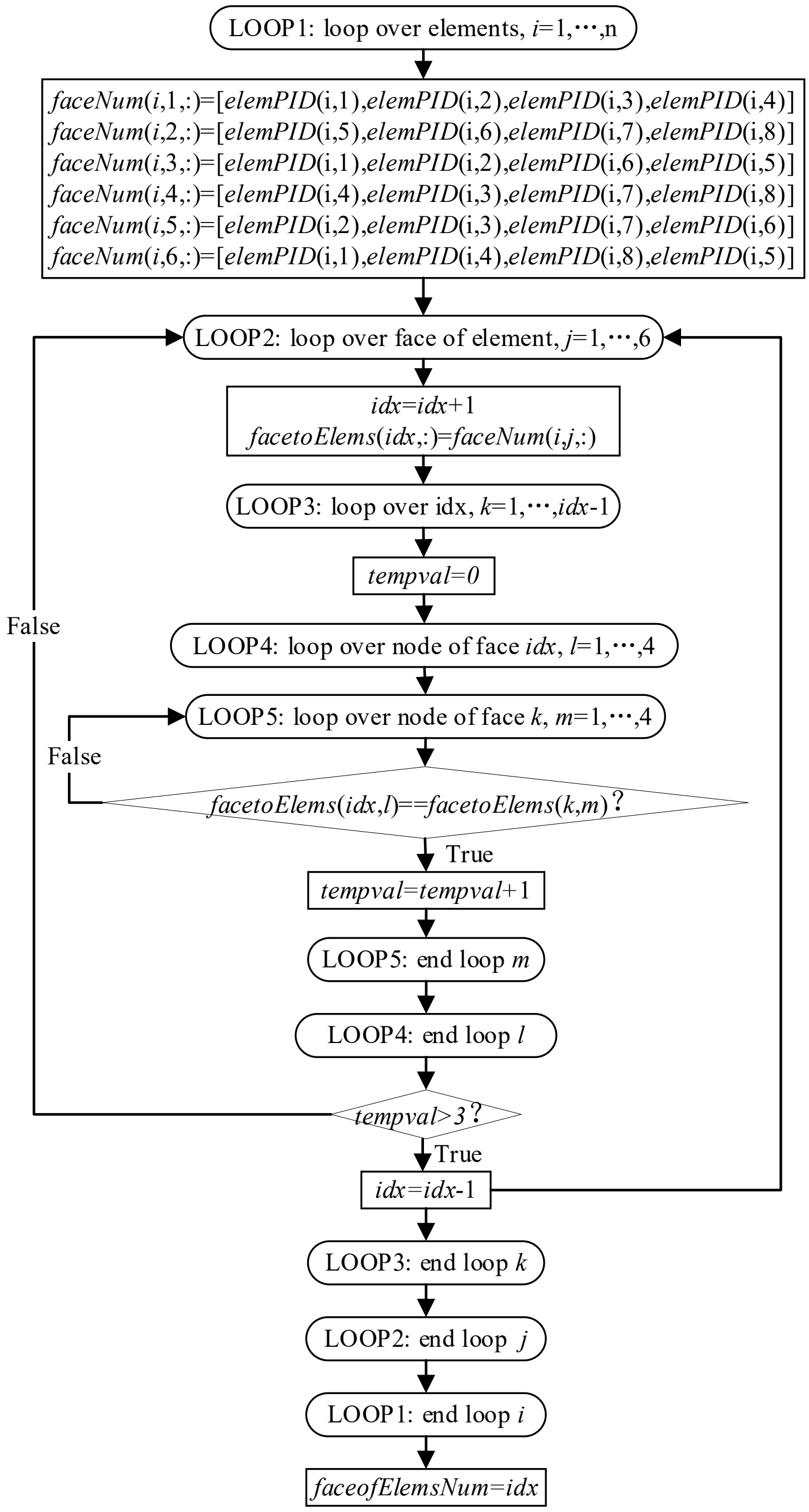

**Fig.5.** Flow chart for recognizing the face of elements.

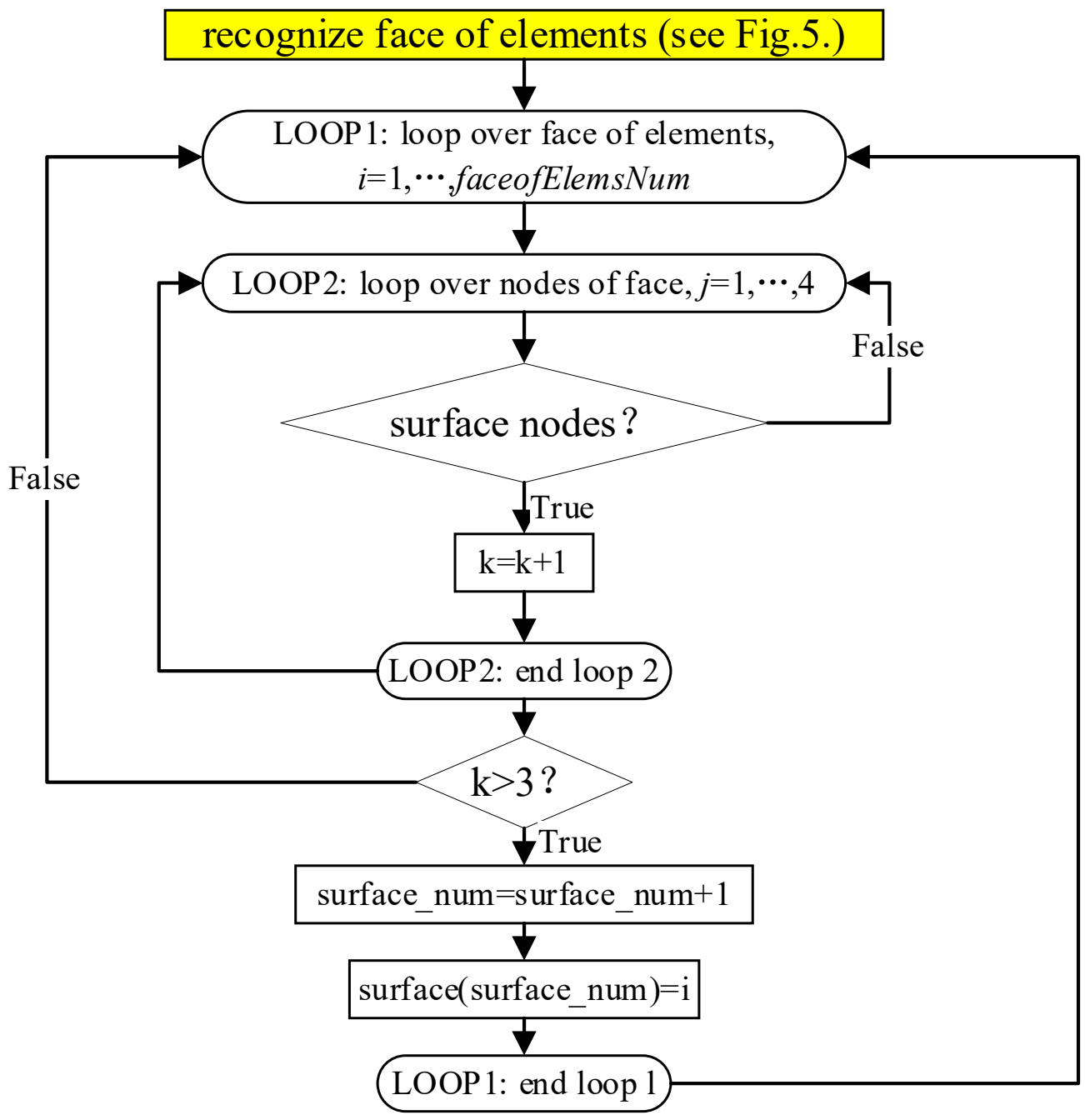

**Fig.6.** Flow chart for recognizing surface.

To construct a contact neighbor list, all the contact particles in the model are searched by establishing a Verlet list. Since all outer element faces have been identified before, a face-based contact neighbor list is then constructed based on the four nodes of each outer surface element face. Finally, a point-to-surface contact search algorithm is used to calculate the positional information of contact pairs, and the penalty function method is then used to calculate the contact force. The detailed computational flow chart is shown in Fig.7.

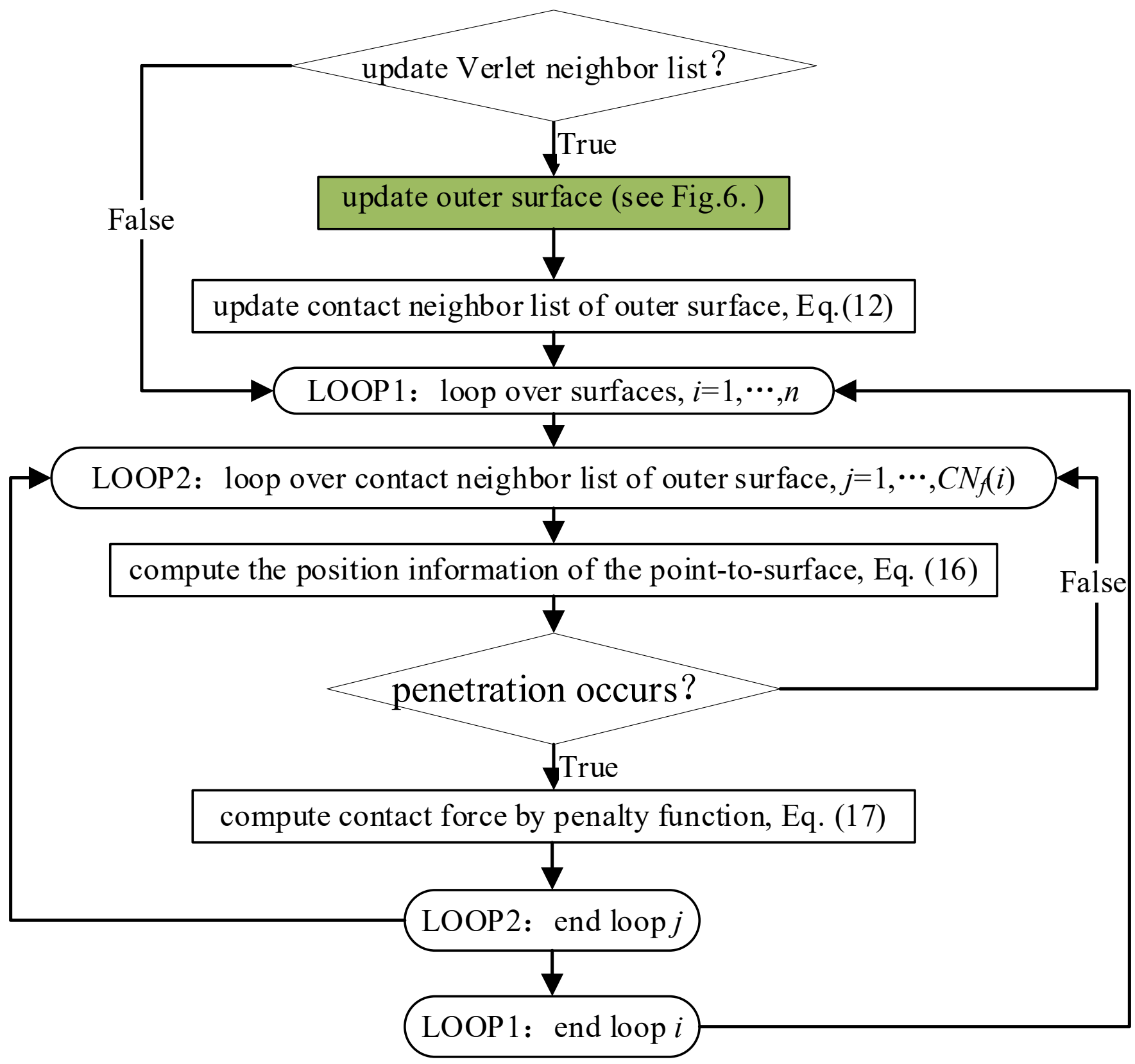

**Fig.7.** Flow chart for calculating the contact force.

### *3.2. Numerical Simulation Procedure*

Here, the numerical implementation of the following numerical examples, shown in Section 4 illustrated, as shown in Fig.8. First, we employ the pre-processing part of the commercial software ANSYS to construct the model and export the node and element information. Subsequently, this information is imported into the PD program and the neighborhood lists is constructed for the particles on the model. The memory allocation and initialization settings are made for the arrays storing the data. At each computational time step, constraints and boundary conditions are applied to the particles in the model, followed by the calculation of the bond force (Eq.(7)) and contact force (Fig.7). The acceleration on each particle is obtained from the bond and contact forces, and the displacements and velocities of the particles are updated and output using the Adaptive Dynamic Relaxation (ADR) [46] method.

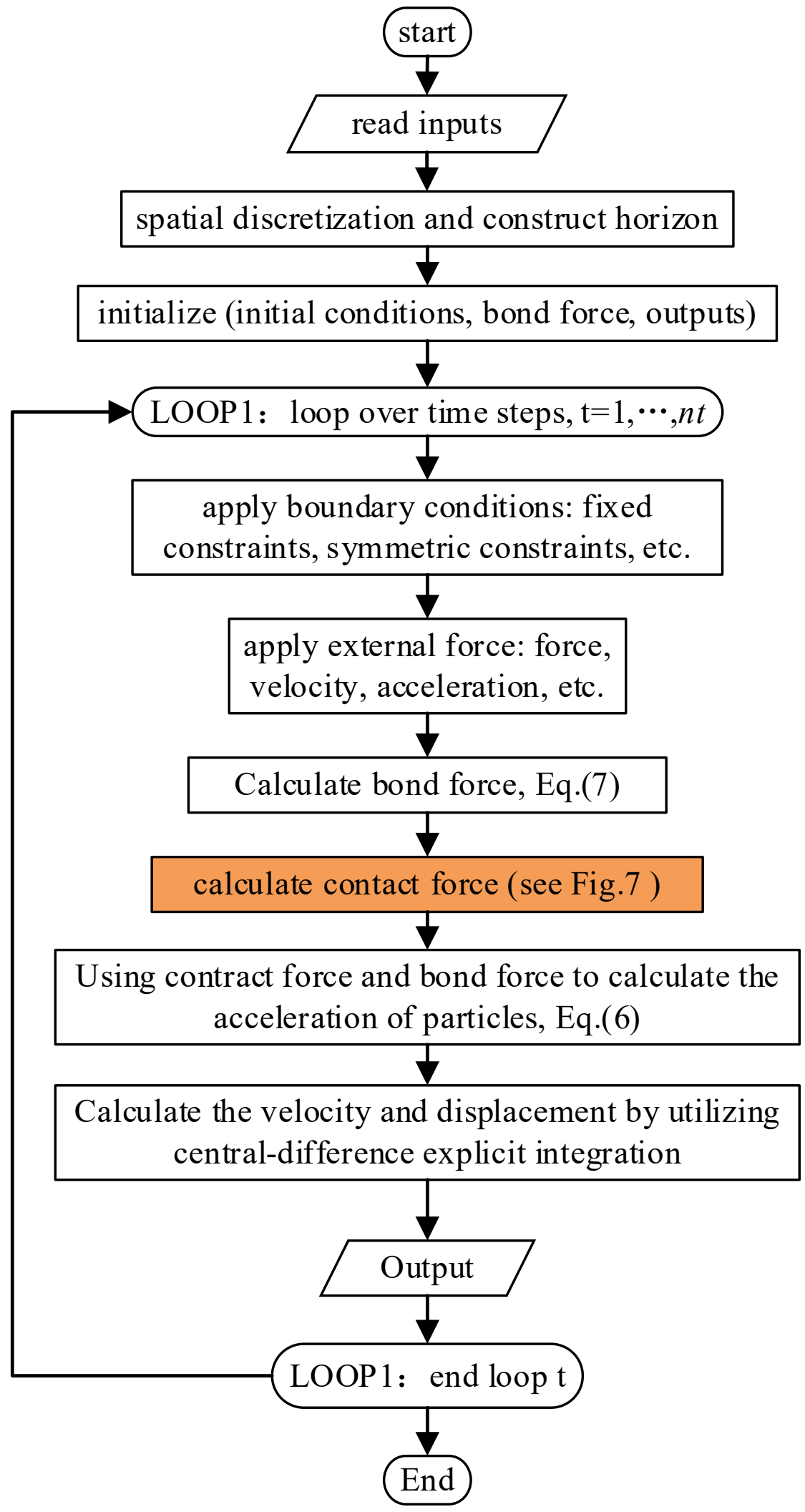

**Fig.8.** Flow chart for the program implementation.

## 4. Verification of the Contact Algorithm in PD Simulations

In this section, we first verify the accuracy of the surface identification algorithm through numerical simulations and determine the optimal eigenvalue threshold based on a systematic parametric analysis. Subsequently, using this contact algorithm, we numerically simulate the contact problem between a rigid punch and an elastic half-space. The effects of grid size, horizon size and timestep on the computational results are also analyzed. The numerical results exhibits well consistencies with theoretical values obtained from Hertz contact theory [47]and the results from other PD classical contact methods. In addition, another two typical examples, i.e., the contact between elastic punch and rigid plane as well as the contact between elastic spheres, are further studied, which demonstrates the accuracy of this algorithm for complex contact problems.

### *4.1. Verification of Surface Identification*

First, it is necessary to determine the range of eigenvalues. As shown in Section 2.2.1, the minimum eigenvalue of $\lambda$ approaches 0 as the particle moves away from the surface, while it tends towards 1 when the particle is inside the body. Here, a basic three-dimensional model is initially constructed to evaluate parameter sensitivity, as shown in Fig. 9. And Fig. 10 displays the recognition performance of external surfaces across varying $\lambda$ values.

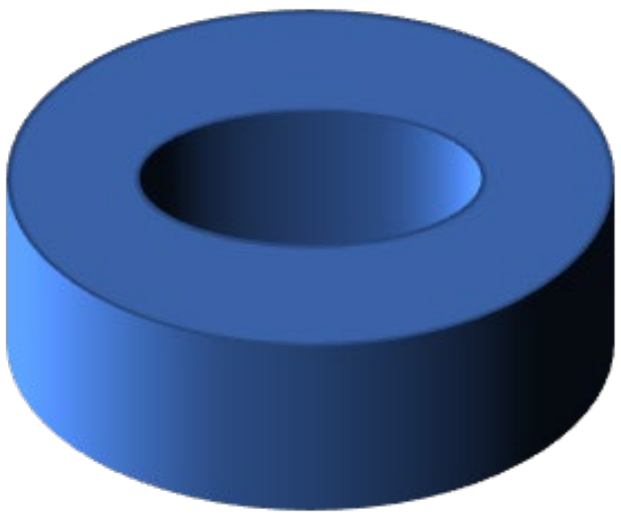

**Fig.9.** The model of hollow cylinder.

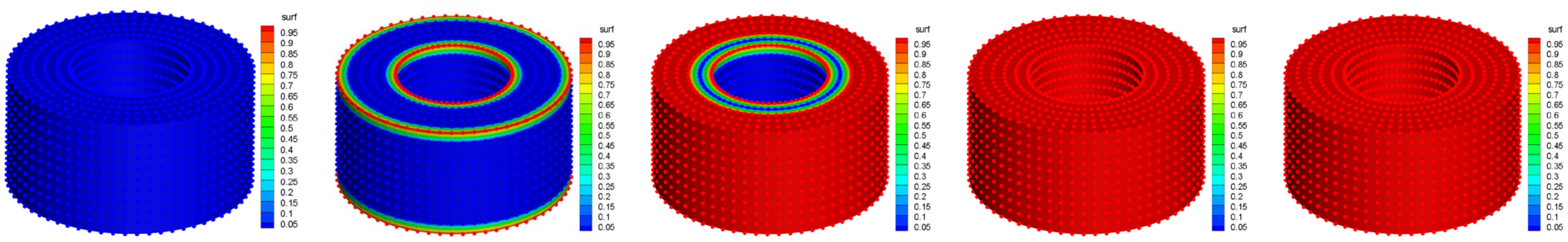

(a) $\lambda = 0.25$ (b) $\lambda = 0.4$ (c) $\lambda = 0.5$ (d) $\lambda = 0.75$ (e) $\lambda = 0.85$

**Fig.10.** The identification effect with different $\lambda$ values.

It suggests that when the value of $\lambda$ is greater than 0.75, all external surfaces can be recognized accurately and completely. Thus, we choose $\lambda$=0.75 in the following calculations, as expressed in the formula below:

$$\begin{cases} \lambda \le 0.75 \Leftrightarrow i \in outside \\ \lambda > 0.75 \Leftrightarrow i \in inside \end{cases} \tag{20}$$

To demonstrate that the optimized $\lambda$ parameter can recognition multi-body systems with various shapes, we perform a heterogeneous morphological test, as shown in Fig.11. The results validate that this algorithm can recognize surface particles of models with various geometries.

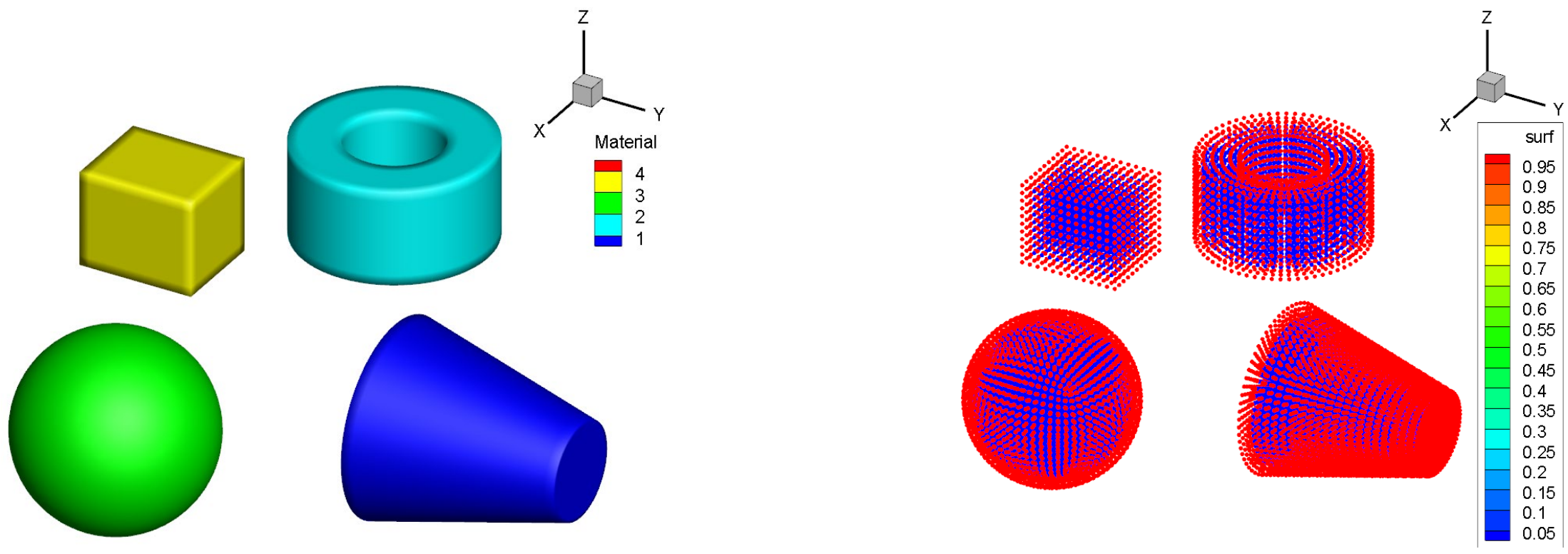

a) multi-body model diagram b) identification of external surface particles

**Fig.11.** Verification the capability of surface identification in multi-body model.

*4.2. Rigid Punch Contact*

To verify the accuracy of the proposed contact algorithm, the contact problem between an elastic half-space and a rigid punch is examined (Fig.12). In the Hertz contact theory [47], for the contact problem between a smooth and frictionless elastic half-space and a rigid punch, the punch has a flat bottom of width $2a$ with sharp corners. When the punch is subjected to a downward external force $P$,the relationship between the pressure and contact position can be expressed as:

$$p(x) = \frac{P}{\pi(a^2 - x^2)^{1/2}} \tag{21}$$

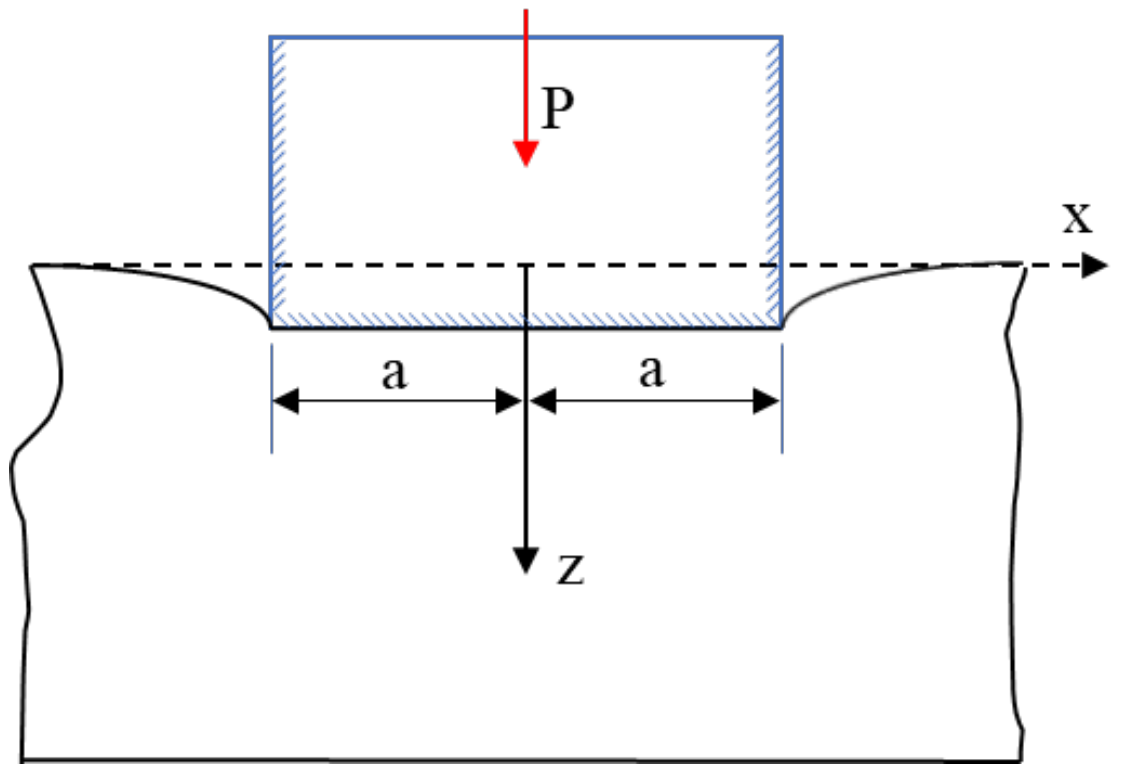

**Fig.12.** Schematic diagram of contact between rigid punch and elastic half-space.

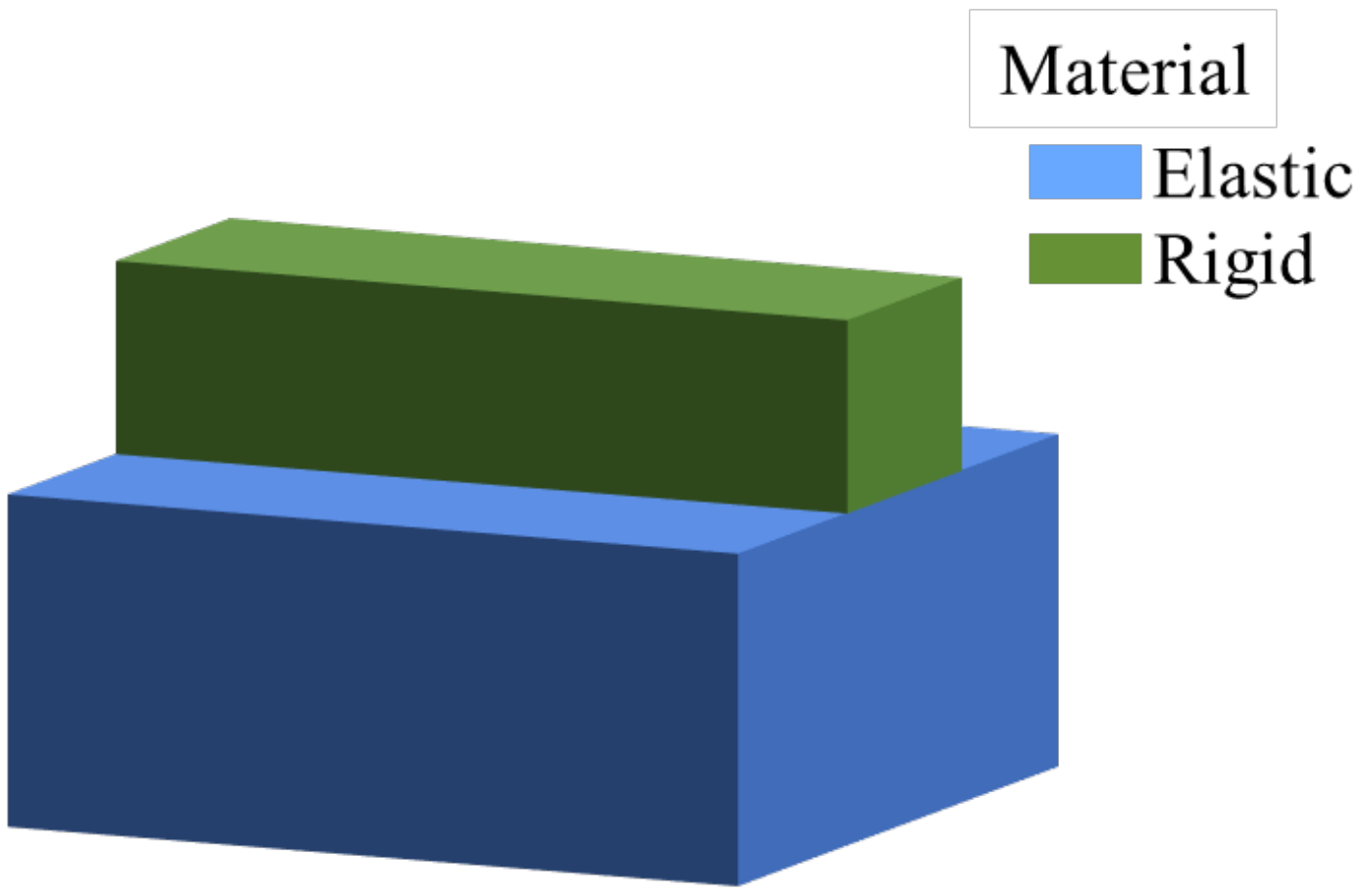

**Fig.13.** Rigid punch and elastic half-space model.

Fig.13 shows the contact model between the elastic half-space and the rigid punch. The punch is considered as a rigid body with dimensions of $4.8\ m \times 1.2\ m \times 1.2\ m$ and density of $1000 \mathrm{kg/m^3}$ . The elastic half-space employs the linear elastic BB-PD model with a density of $1000 \mathrm{kg/m^3}$, an elastic modulus of 1GPa, and a Poisson's ratio of 0.25. In order to facilitate the calculation of the contact force, the material properties of rigid punch are set to be the same as those of the elastic half-space, although it does not deform. The grid size of the model is $\Delta$=0.08m. The horizon size of the model is chosen to be $8th$ neighborhood ($m = 3$). The convergency validation will be illustrated in Section 4.3. A vertically downward external load $P = 1.66\times10^8 N/m$ is applied to the rigid punch.

During the loading process, the contact force is in equilibrium with the external load $P$ when the punch interacts with the elastic half-space and causes deformations. In this equilibrium state, the distribution of normal contact pressure on the contact surface aligns with the mathematical form of Eq.(21). Then we compare the calculated contact forces with theoretical values (see Eq.(21))and those obtained from the rigid body contact collision algorithm[23] and the penalty-function based point-to-point contact algorithm[41], are shown in Fig.14.

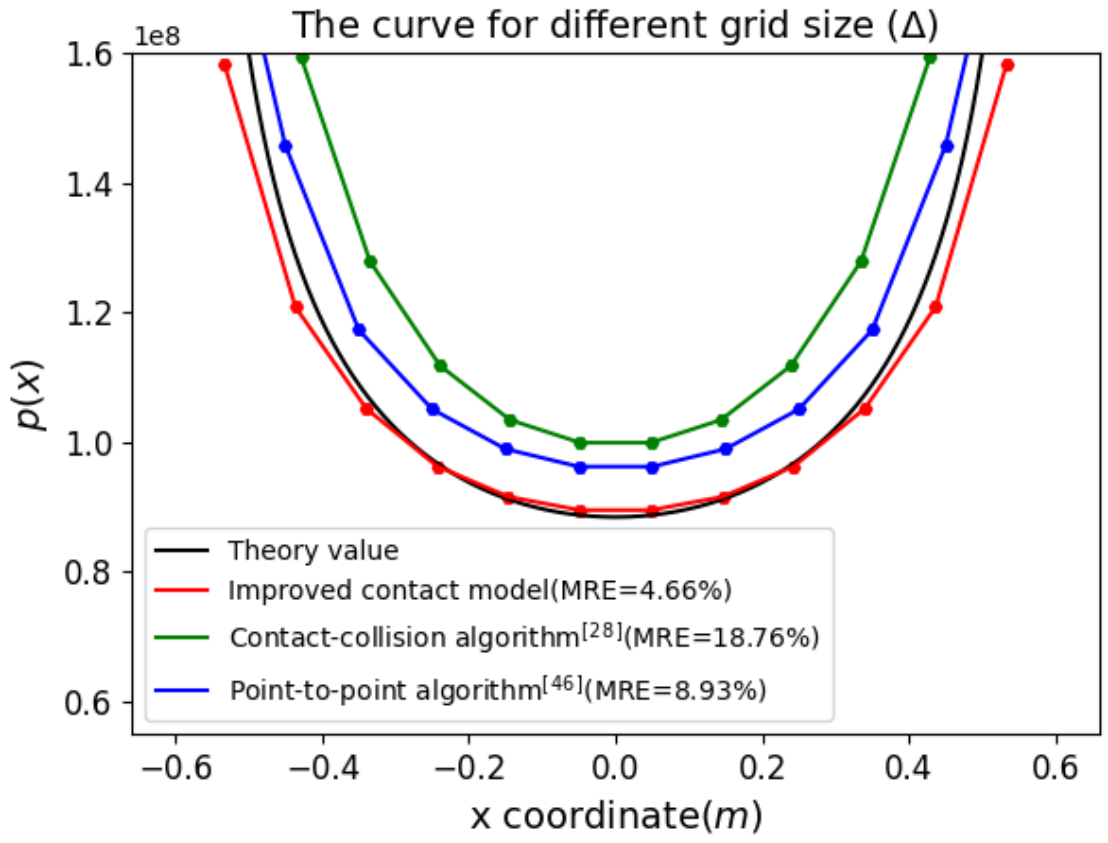

**Fig.14.** Contact force curve with position.

As shown in Fig. 14, the mean relative error (MRE) between the contact force calculated by the proposed contact algorithm and the theoretical values is only 4.66%. While the contact forces obtained from the rigid body contact collision algorithm and the point-to-point contact algorithms exhibit large deviations of 18.76% and 8.93%, respectively, from theoretical results. Therefore, our proposed contact method can estimate the contact force accurately.

### *4.3. Convergence Analysis*

In this Section, we perform a convergence analysis of the contact problem between a rigid punch and an elastic half-space, as shown in Section 4.2, specifically examining $m$-convergence, $\delta$-convergence and $\Delta t$-convergence.

For $\delta$-convergence, the parameter *m* remains constant, while $\delta$ varies based on $\delta = m\Delta$. Here, we employ the $8th$ neighborhood of model $(m = 3)$ with four grid sizes: Δ=0.12m, 0.1m, 0.08m and 0.06m. The Z-axis contact forces at these grid sizes are shown in Fig.15. As the mesh size decreases, the MRE is only 3.91% when the mesh is 0.08m, while the error of the simulation results for the mesh size of 0.06m is only 1.02%, which suggests that the mesh size reaches convergence at 0.08m. Thus, in the present study, we select 0.08m as mesh size.

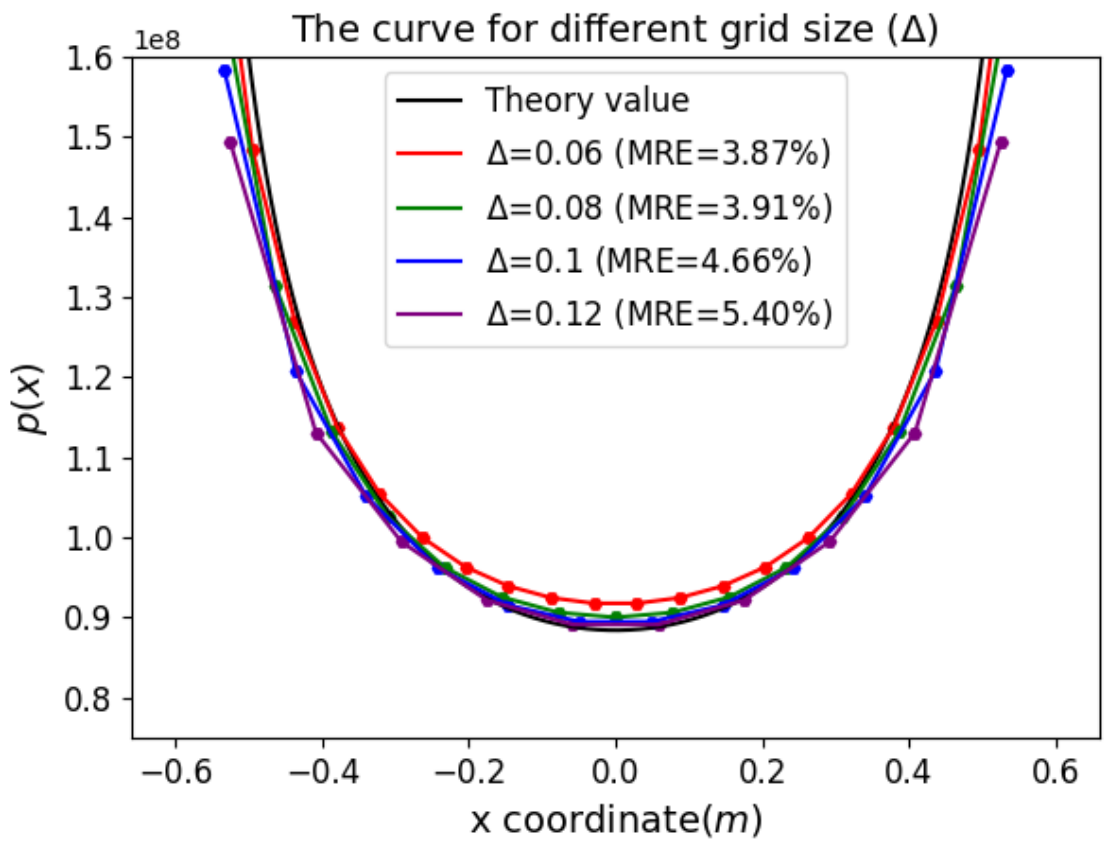

**Fig.15.** Contact force curve with position for different grid sizes.

For $m$-convergence, the value of $\Delta$ is set as a constant value of 0.08. And four values of the horizon sizes are considered: the $7th$ neighborhood $(m = \sqrt{8})$, $8th$ neighborhood $(m = 3)$, $9th$ neighborhood $(m = \sqrt{10})$, and $10th$ neighborhood $(m = \sqrt{13})$. Fig.16 shows the contact force distribution curves with different neighborhood horizon sizes. The results suggest that when *m* reaches 3, i.e., the 8th nearest neighbor, the contact force exhibits the best agreement with the theoretical value, and the MRE of the curve is only 3.91%.

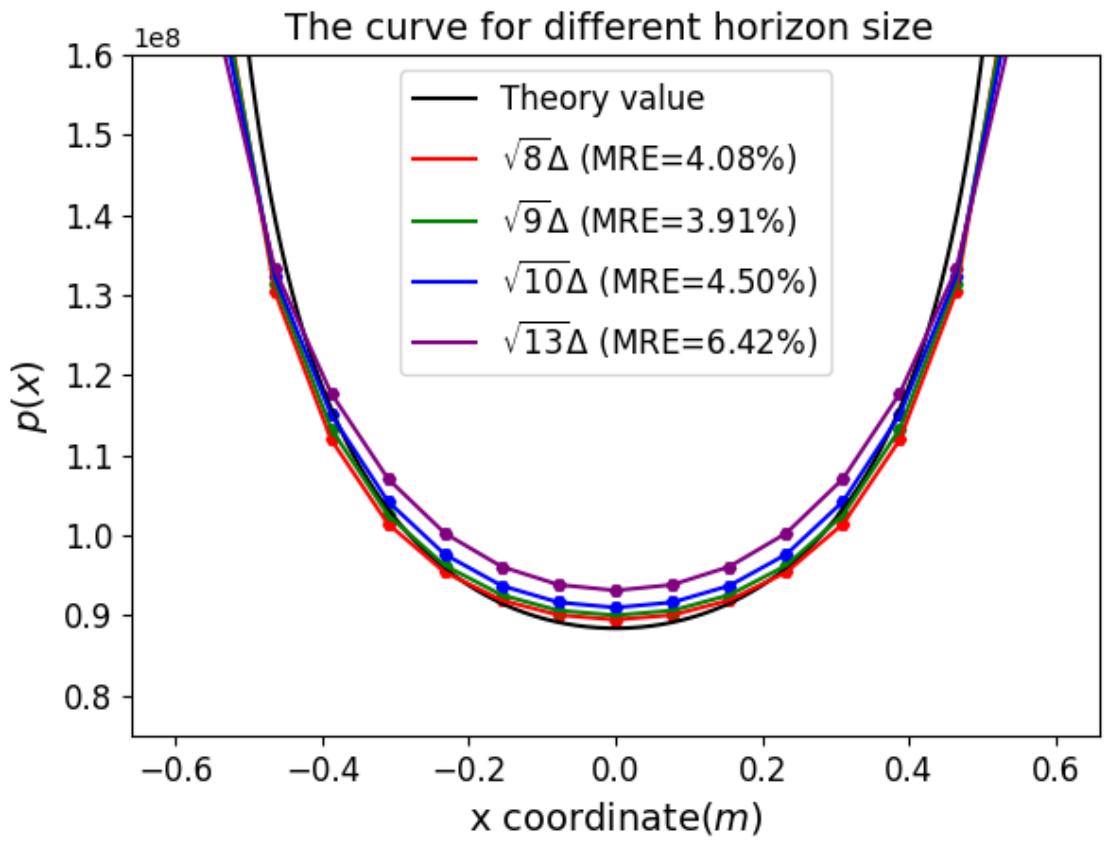

**Fig.16.** Contact force curve with position for different horizon sizes.

For $\Delta t$-convergence, the grid size and the horizon size of the model are set as 0.08m and 3, respectively. Here, three different timesteps are considered: $\Delta t = 1\times10^{-5}$s, $5\times10^{-5}$s, and $1\times10^{-6}$s. The relationship between the contact force and its position at different timesteps are shown in Fig.17. The findings demonstrate that the magnitude of contact force remains constant at the same position across different time steps. Thus, this contact algorithm is independent of the time step, and is exclusively material-dependent.

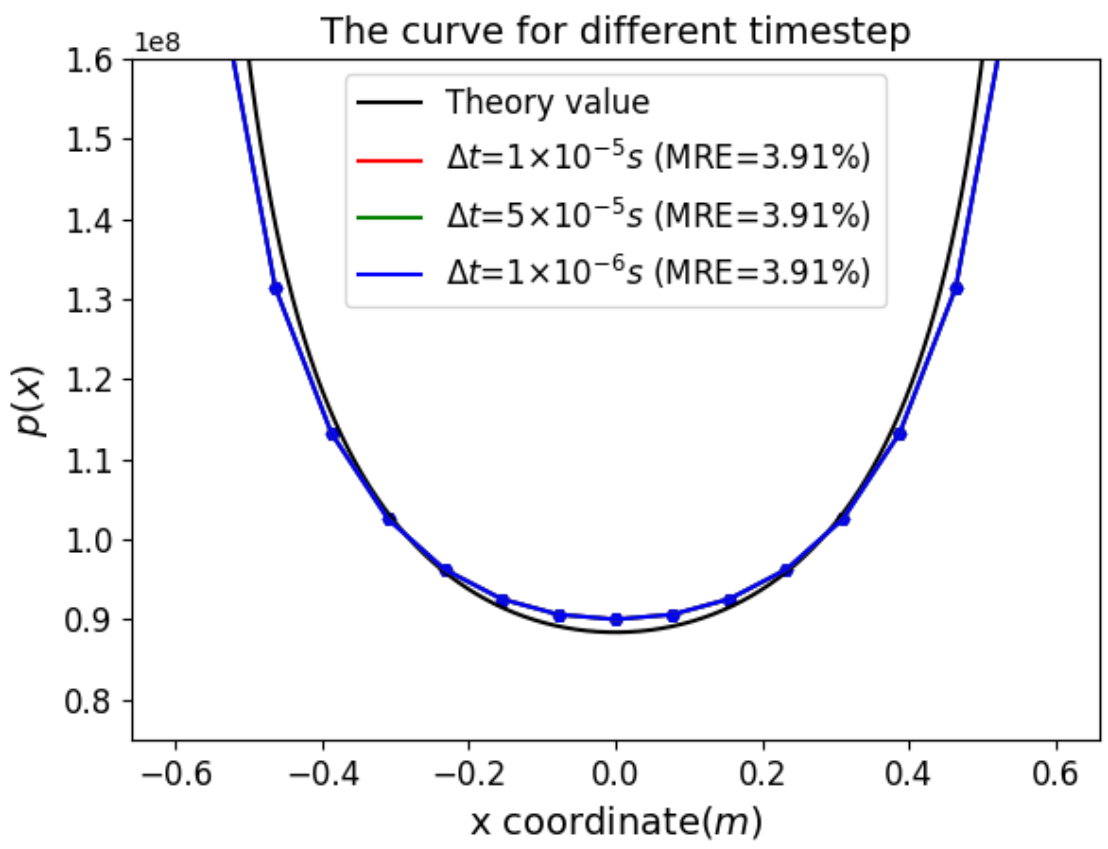

**Fig.17.** Contact force curve with position for different time steps.

## *4.4. Elastic Roller Contact*

Next, the contact algorithm is employed to address the contact problem between an elastic roller and a rigid plane. The roller is subjected to a downward external force $P$, which contacts the plane and forms a rectangular surface with a contact half-width of $a$, as shown in Fig.18. The pressure on the contact surface between the roller and the plane exhibits a parabolic relationship with the contact half-width, as shown in Fig.19, it is expressed as

$$p(x) = p_0[1 - (\frac{x}{a})^2]^{1/2} \tag{22}$$

where the expressions of contact radius $a$, maximum pressure $p_0$ and average pressure $p_m$ are respectively:

$$a = \left(\frac{4}{\pi}\frac{PR}{E'}\right)^{1/2} \tag{23}$$

$$p_0 = \frac{2P}{\pi a} \tag{24}$$

$$p_m = \frac{P}{2a} = \frac{\pi}{4}p_0 \tag{25}$$

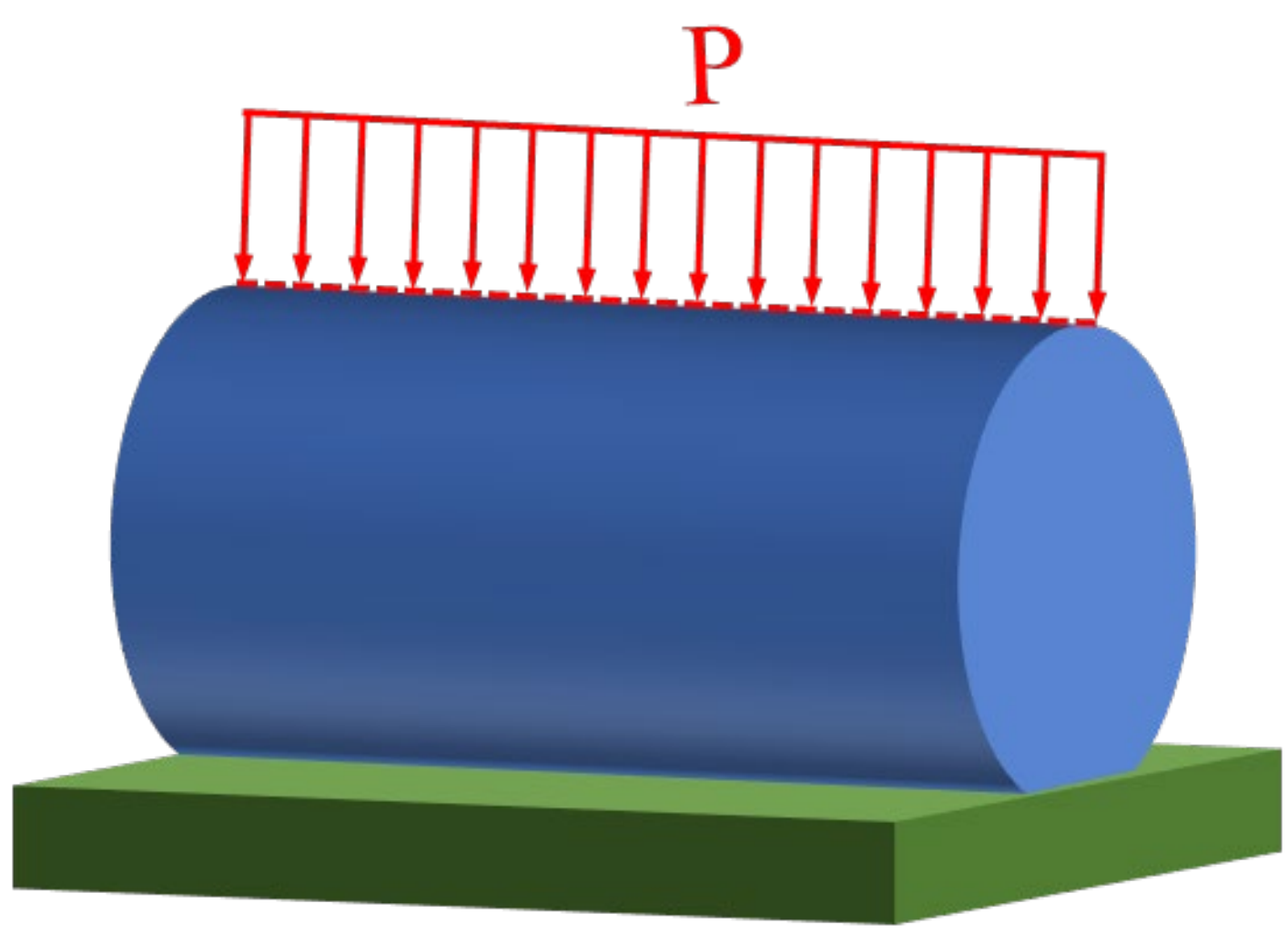

**Fig.18.** Diagram of elastic roller contact.

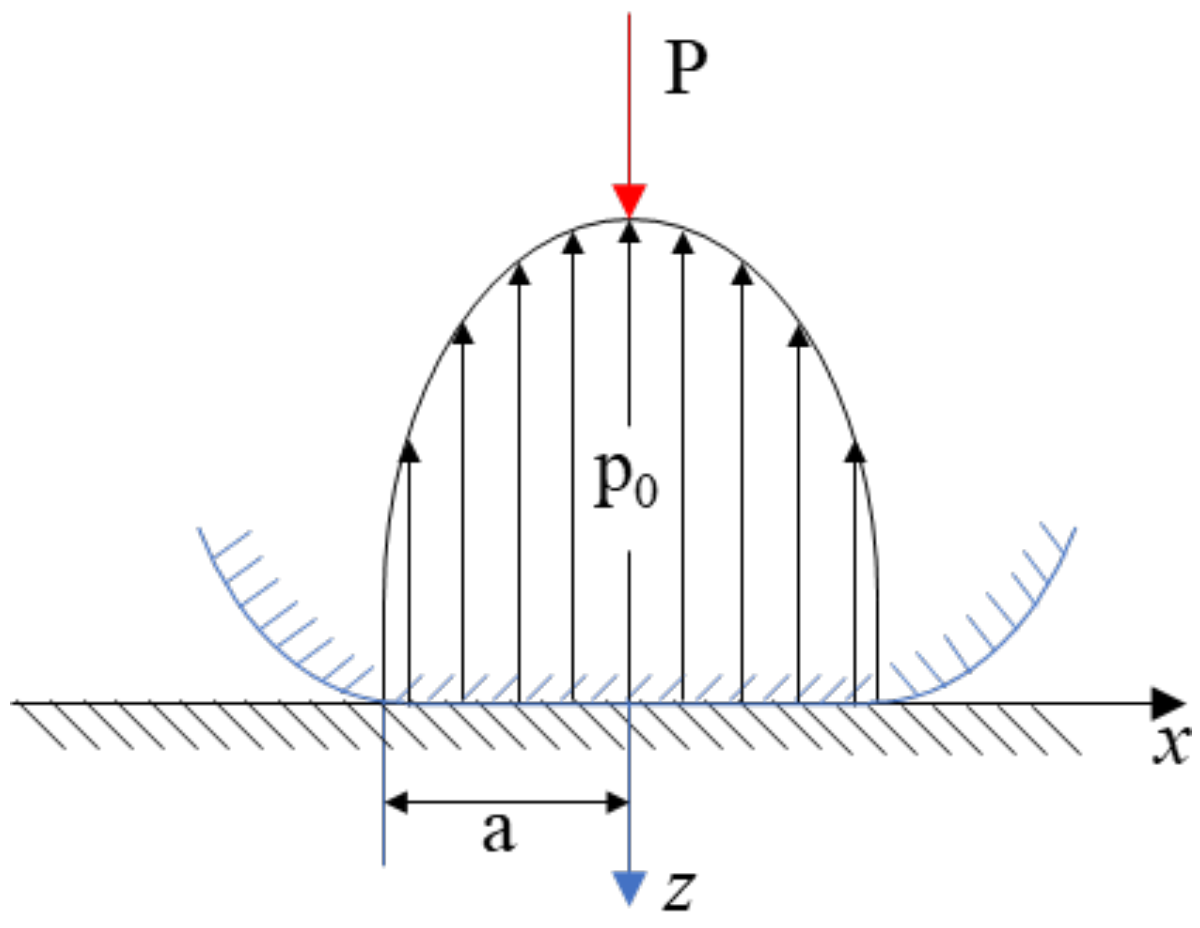

**Fig.19.** Schematic diagram of contact pressure distribution of elastic rollers.

As shown in Fig.18, the roller has a radius of 1m, a length of 4m, a density of $1000 kg/m^3$, and an elastic modulus of 1GPa. The material properties of the rigid body are the same as those of the elastic roller. The roller adopts the linear elastic BB-PD model, and the plane is rigid. Based on the convergency results, the grid size of the model is Δ=0.08m, and the horizon size is chosen to be the $8th$ neighborhood of model ($m = 3$). A downward external force $P = 8\times10^8 N/m$ is applied to the roller, when the contact force and external force $P$ reach equilibrium. The distribution of normal phase pressure satisfies the expression of Eq. (22).

In Fig.19, when the elastic roller is subjected to a downward linear load, it often contacts with the rigid plane and forms a rectangular contact area. According to Eq. (23) and Eq. (24), the theoretical values of the contact half-width and the maximum contact stress are 0.69099m and $1.84264\times10^8$Pa, respectively. As shown in Fig.20, the MRE of contact force between contact algorithm and theoretical value is only 6.81%, while the contact half-width obtained from fitting against the simulation results are in general agreement with the theoretical results.

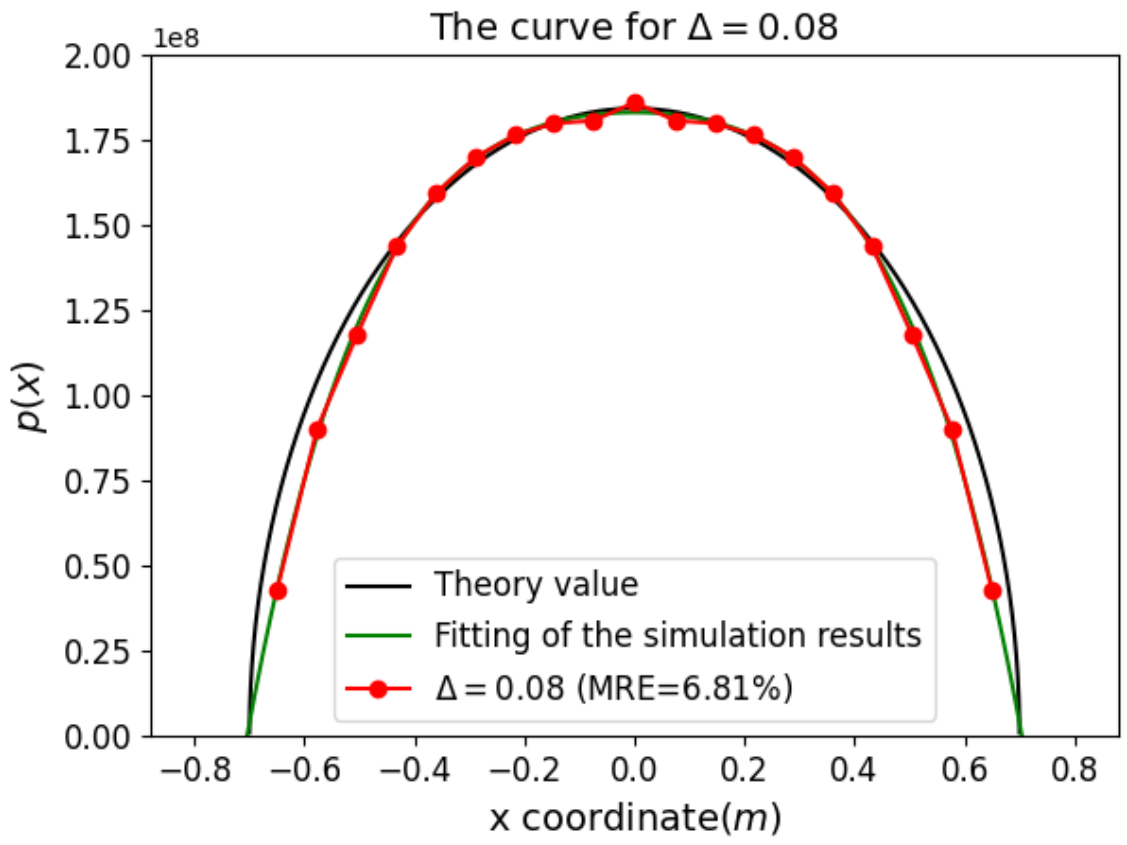

**Fig.20.** Contact force curve with position.

### *4.5. Elastic Sphere Contact*

Here, we examine the contact problem between two identical elastic spheres. The two spheres are in contact with each other under the action of an external force P, forming a contact circle of radius a, as shown in Fig.21. The pressure on the contact surface is parabolically related to the contact radius[47], expressed as

$$p(x) = p_0[1 - (\frac{x}{a})^2]^{1/2} \tag{26}$$

Among them, the expressions for contact radius $a$, maximum pressure $p_0$, and average pressure $p_m$ are respectively.

$$a = \left(\frac{3}{4}\frac{PR}{E'}\right)^{1/3} \tag{27}$$

$$p_0 = \frac{3}{2}\frac{P}{\pi a^2} \tag{28}$$

$$p_m = \frac{P}{\pi a^2} = \frac{2}{3}p_0 \tag{29}$$

$$E' = \left(\frac{1-v_1^2}{E_1} + \frac{1-v_2^2}{E_2}\right)^{-1} \tag{30}$$

$$R = \left(\frac{1}{R_1} + \frac{1}{R_2}\right)^{-1} \tag{31}$$

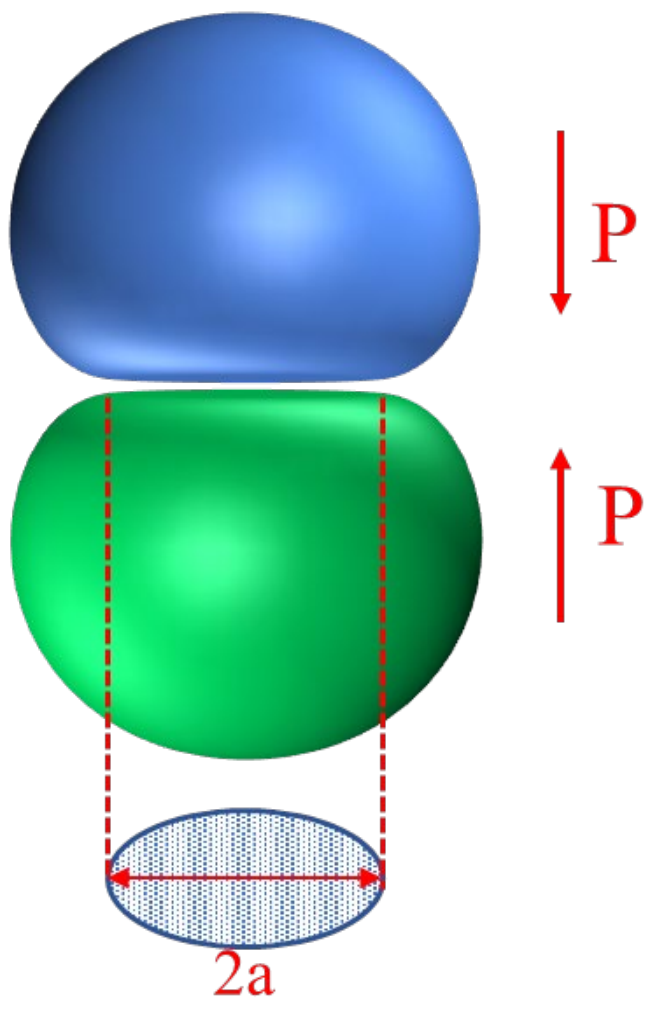

**Fig.21.** Diagram of elastic sphere contact.

The contact model of two elastic sphere is constructed with a radius of 1m. The density, elastic modulus, and Poisson's ratio of the spheres are $1000\text{kg/m}^3$, $1\text{GPa}$, and $0.25$, respectively. The spheres use the linear elastic BB-PD model. The grid size and horizon size of the model are set as 0.08m and the $8th$ neighborhood of model ($m = 3$), respectively, after the convergency tests. A paired external force $P$ with a magnitude of $1\times10^8 N$ is simultaneously applied to the two spheres to bring the spheres into positive contact.

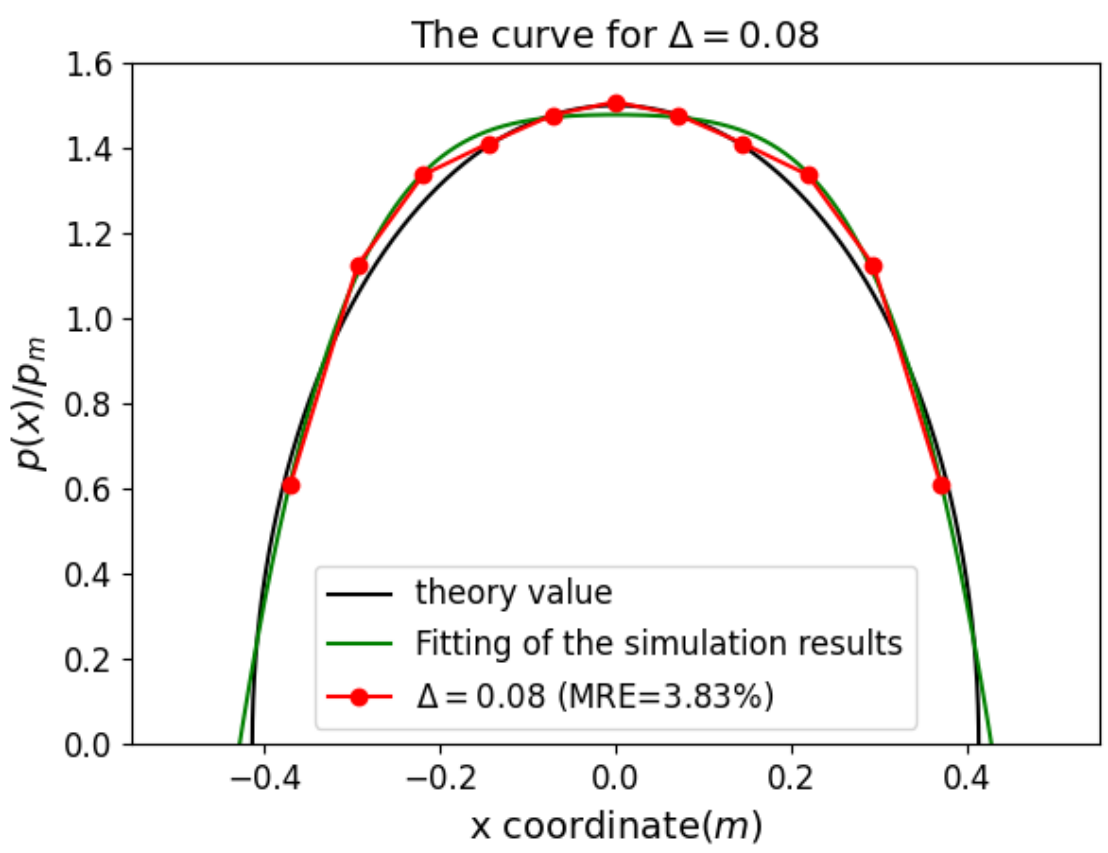

**Fig.22.** Contact force curve with position.

As shown in Fig.21, when the two spheres are subjected to an external force in opposite directions,

contact will occur between the spheres and a circular contact zone will be formed. Based on Eq.(27) and Eq. (28), the theoretical values of the contact radius and the maximum contact stress are 0.41274m and $2.80276\times10^{8}$Pa, respectively. The values of $p(x)/p_m$ of the particles on the sphere surface with the position curve are shown in Fig.22. Compared with theoretical value, the MRE of the contact algorithm is only 3.83%, demonstrating the accuracy of our proposed contact force algorithm. Meanwhile, the results of the simulation are fitted for the simulated results, and the contact radius obtained from the fitting is about 0.428m with the relative error compared to the theoretical value by about 3.69%.

## 5. Conclusion

In this study, we propose an improved point-to-surface contact algorithm, which can automatically identify contact surfaces and accurately predict contact forces. First, this algorithm employs the eigenvalue method with a surface recognition threshold of $\lambda$=0.75 to detect contact surfaces. A global search is then performed using a Verlet list to efficiently capture all potential contact pairs. Next, the precise contact locations of all potential contact pairs are subsequently determined through a point-to-surface contact search, and the corresponding contact forces are estimated using the penalty function method. Finally, several numerical simulations are performed to demonstrate the accuracy of this proposed contact algorithm, including contact interactions between an elastic body and a rigid plane as well as two elastic bodies. The calculated contact forces exhibit well consistent with those from Hertz contact theory, validating the accuracy of this algorithm. Notably, the estimated contact forces also demonstrate better agreement with Hertz theoretical solutions compared to those obtained from other PD contact models. The present work contributes a more direct and accurate contact algorithm for PD simulations, offering a reliable solution for addressing multi-body contact and collision challenges in practical engineering applications.

**Acknowledgments**

J. Li was supported from the National Natural Science Foundation of China (No. U2441215). L. Liu, X Lai, and H Mei were supported from the National Natural Science Foundation of China (No. 52494933). X. Lai was supported from the Open Fund of the Hubei Longzhong Laboratory (No. 2024KF-30), the Open foundation of the Hubei Key Laboratory of Theory and Application of Advanced Materials Mechanics and the Open Foundation of Hubei Key Laboratory of Engineering Structural Analysis and Safety Assessment.